\documentclass[aps,floatfix,nofootinbib]{revtex4}
\usepackage{latexsym}
\usepackage{mathptm}
\usepackage{amsmath}
\usepackage{amssymb}
\usepackage{graphicx}
\def\beq{\begin{equation}}
\def\eeq{\end{equation}}
\def\beqn{\begin{eqnarray}}
\def\eeqn{\end{eqnarray}}
\newcommand{\f}{\begin{equation}}
\newcommand{\ff}{\end{equation}}

\begin{document}

\title{OPERA neutrinos and deformed special relativity}
\author{Giovanni Amelino-Camelia$^a$, Laurent Freidel$^b$, Jerzy Kowalski-Glikman$^c$, Lee Smolin$^b$
\\
$^a$Dipartimento di Fisica, Universit\`a ``La Sapienza" and Sez.~Roma1 INFN, P.le A. Moro 2, 00185 Roma, Italy\\
$^b$Perimeter Institute for Theoretical Physics, 31 Caroline Street North, Waterloo, Ontario N2J 2Y5, Canada\\
$^c$Institute for Theoretical Physics, University of Wroclaw,  Pl. Maxa Borna 9, 50-204 Wroclaw, Poland
}

\begin{abstract}
In a recent study, Cohen and Glashow  argue that superluminal neutrinos
of the type
recently reported by OPERA should be affected
by anomalous Cherenkov-like processes.  This causes them to loose much of
 their energy before reaching the OPERA detectors. Related concerns were reported
 also by Gonzalez-Mestres, Bi et al, and  Cowsik et al,
 who argued that pions cannot decay to superluminal neutrinos over
 part of the energy range studied by OPERA.
We observe here that these
arguments are set within a framework in which
Lorentz symmetry is broken, by the presence of a preferred frame.
We further show that these anomalous processes are forbidden
if Lorentz symmetry is instead ``deformed", preserving the relativity of inertial frames.  These deformations add non-linear terms to
energy momentum relations, conservation
laws and Lorentz transformations in a way that is consistent with the relativity of inertial observers.
\end{abstract}

\maketitle


The OPERA collaboration recently reported~\cite{opera} evidence of
superluminal behavior for $\mu$ neutrinos with energies of a few
tens of GeVs: ${v} - 1 \simeq 2.4 \cdot 10^{-5}$, with a
significance of six standard deviations (we use units such that the
speed of light is $c=1$).

As usual in science, when a particularly striking experimental
result is first reported, the most likely hypothesis is that some
unknown bias or source of uncertainty affects these OPERA data.
However, the result would be of such potential importance that it
behoves us to at least investigate what is the second most likely
hypothesis that is consistent with all the relevant data, including
OPERA, and gives a reasonable path for physics to go ahead.  In the
event that the OPERA data is right, we must, by the development of
such an hypothesis, come to understand special relativity as an
approximation to  deeper physics.  This is of no small importance
because a particular hypothesis that can explain all the relevant
data can serve to predict other experiments.  These predictions, if
confirmed, would strengthen the case that there really are
departures from special relativity.

However, a recent letter by Cohen and Glashow~\cite{glashowOPERA}
 appears to indicate that there cannot
be such a second hypothesis, since it claims that the OPERA results
are self-contradictory.
Cohen and Glashow argue that neutrinos in OPERA's CNGS beam,
if superluminal  as described in Ref.~\cite{opera},
would loose much of their energy
 via  Cherenkov-like processes on their way from CERN to Gran Sasso.  They
 then could not be detected with energy in excess of $~12 GeV$, contrary
 to what is also reported~\cite{opera} by OPERA.
And a similar message is contained in studies by Gonzalez-Mestres~\cite{gonzaOPERA},
 by Bi et al~\cite{Bi}, and by Cowsik et al~\cite{nussinov},
which argued
 that the reported superluminality of $\mu$ neutrinos would even prevent,
 over part of the energy range studied by OPERA,
 the  pion-decay processes at CERN,
 which are partly responsible for the flux of neutrinos reaching Gran Sasso.
 If these arguments are correct, then it seems unavoidable that the OPERA results
 are mistaken.

But, as we show here, there is a loophole\footnote{Cohen and Glashow mention the possibility of
a different loophole in which electrons and photons share with neutrinos
``unconventional dispersion relations... such that, in the energy domain of the
OPERA experiment, these particles travel with a common
velocity"~\cite{glashowOPERA}.
This would indeed render the OPERA data no longer self-contradictory;
however, our understanding is that this is in contradiction with established upper bounds on
the speed of multi-GeV electrons\cite{N} and of multi-GeV photons\cite{N+1}.}
in the arguments reported
in Refs.~\cite{glashowOPERA,gonzaOPERA,Bi,nussinov}, due to an implicit
assumption they make about the fate of Lorentz invariance.

As already established in Ref.~\cite{whataboutopera},
the possibility that these neutrinos are tachyons, consistently with
special relativity, must be excluded, if we take into account other available
data on the possible dependence of the neutrino maximal
speed with energy.
Any attempt to interpret the OPERA anomaly as actual evidence of neutrinos
with superluminal behavior must therefore accommodate departures from
Lorentz symmetry.

Before we begin such an analysis it is crucial to recall that Lorentz invariance can be superseded in two ways.
It can be ``broken"~\cite{kosteSMEfirst,colglafirst,grbgac}
 in the sense that there is a preferred frame of
reference. Or it can be ``deformed"~\cite{gacdsr,jurekdsr,leedsr},
so that
the principle of relativity of inertial frames is
preserved, but the action of Lorentz transformations on physical states
is deformed.
In such a deformation of special relativity the energy-momentum
relations, conservation laws, and Poincar\'e transformations are all
modified by non-linear terms, which are mutually consistent so that
the relativity of inertial frames is preserved.

However, as we show below, the
Cherenkov-like processes considered in Ref.~\cite{glashowOPERA} and
the features of pion decay discussed
in Refs.~\cite{gonzaOPERA,Bi,nussinov},
necessarily assume that Lorentz invariance is broken. Hence, their
claims that the OPERA results contradict known physics are only
valid in a framework in which Lorentz invariance is broken by the
presence of a preferred frame.

Furthermore, as we also show below, if one instead interprets the
OPERA results within a framework in which the relativity of inertial
frames is unbroken, but the laws of special relativistic kinematics
are modified, then the objections of  Refs.~\cite{glashowOPERA,gonzaOPERA,Bi,nussinov}
are not valid. Superluminal neutrinos may be
produced by pion decay over the whole range of OPERA energies and,
once produced, are stable against loss of energy through
Cherenkov-like processes. These results,  combined with the ones
previously reported in Refs.~\cite{glashowOPERA,gonzaOPERA,Bi,nussinov},
imply that any attempt to interpret the OPERA anomaly in terms of
superluminal neutrinos should favor interpretations  in which Lorentz
symmetry is deformed but not broken.


We now will contrast the interpretation of the OPERA experiment in
two scenarios, in which the relativity of inertial frames is
respectively, broken and deformed.  These two frameworks have in
common that the energy-momentum relation is modified by non-linear
terms governed by a characteristic mass scale $M_*$. We start with a
general form of dispersion relation
\begin{equation}
 E^2\, f^2(E,p;\mu^2) - p^2\, g^2(E,p;\mu^2) = m^2
 \label{dispmod}
 \end{equation}
with $$f(E,p;\mu^2)= 1 + \phi(E,p;\mu^2), \quad g(E,p;\mu^2)= 1 +
\gamma(E,p;\mu^2)$$ where $\phi$, $\gamma$ are functions, much
smaller than $1$ for in the relevant regime, codifying departures
from special relativity, $m$ is the particle rest mass and $\mu^2$
is a deformation scale that,  {\it  crucially for what follows, can be
different for different particles.}

Assuming  (will return to this point in the closing remarks) that
the observer adopts spacetime coordinates conjugate to the
momentum-space variables $E,p$ one easily finds that as a
consequence of (\ref{dispmod}) the speed of a particle depends on
energy as follows
\begin{equation}
v(E) = \frac{dE}{dp} = \frac{\bar{p}}E -
\frac{d}{dE}\left(\bar{p}\,\phi(E,\bar{p};\mu^2)\right)
+\frac{\bar{p}^2}{E^2}\,\frac{d}{dE}\left(\bar{p}\,\gamma(E,\bar{p};\mu^2)\right)
 \label{velmod}
 \end{equation}
where $\bar{p}\equiv \sqrt{E^2-m^2}$ and we worked at leading order
in $\phi$, $\gamma$.

In the framework with a preferred frame these equations hold only in
that special frame. In this case it is legitimate to  hypothesize
that energy momentum conservation remains linear in all frames of
reference.  However the energy-momentum relations (\ref{dispmod})
alter their form~\cite{gacdsr} in other than the preferred frame.
This, as we will show below, is the context in which Cohen and
Glashow, Gonzalez-Mestres, Bi et al, and  Cowsik et al,
 derive their conclusions.

However, if we want the relativity of inertial frames to be
preserved then we want (\ref{velmod}) and (\ref{dispmod}) to hold in
all inertial frames. This is possible~\cite{gacdsr,jurekdsr,leedsr}
 if the action of the generators of
Lorentz boosts ${\cal N}_j$ is modified in such a way that (without
modifying the standard algebra of Lorentz generators)
 \f \big[{\cal N}^{(\mu)}_j , E^2\, f^2(E,p;\mu^2) - p^2\, g^2(E,p;\mu^2)
\big]=0 \label{ee} \ff The next step is to realize that if the
action of the boosts is modified, then the energy-momentum
conservation laws must modified as well, so that they stay covariant
\cite{gacdsr,leedsr,judesvisser}.

The key differences between the broken-Lorentz and the
deformed-Lorentz frameworks are two: in the broken case the
conservation laws remain the standard linear laws in every frame but
the modified energy-momentum relation (\ref{dispmod}) hold only in
the preferred frame. However in the deformed Lorentz framework the
modified energy-momentum relations hold in all inertial frames and
the conservation laws become non-linear as well.

The arguments of Refs.~\cite{glashowOPERA,gonzaOPERA,Bi,nussinov}
adopt modified dispersion relations but relied on unmodified
energy-momentum conservation, so they are situated in the framework
of broken Lorentz invariance.

One consequence of the presence of a preferred frame is that
processes with only one incoming particle can be allowed or
forbidden, depending on whether the energy of the incoming particle
is above or below a certain threshold value. This cannot happen in
relativistic frameworks, even in cases in which the Lorentz
transformations have been deformed \cite{gacnewjourn,sethmajor}.

This is illustrated in the example of the process
\f
\nu_\mu \rightarrow \nu_\mu  + e^+  + e^-
\ff
which is the main focus of Ref.~\cite{glashowOPERA}.
Cohen and Glashow find that this process is forbidden
at low energies, but above a certain threshold energy $E_{thresh}$,
in the preferred frame,
the superluminality of the neutrino becomes sufficient to render this
process  allowed.

To see that this requires a preferred frame note that the threshold
in question is not Lorentz invariant as is the case of thresholds
found in special relativistic kinematics.  Consider the case in
which the preferred frame observer Alice sees neutrinos with energy
just over the threshold,
 $E_{Alice} = E_{thresh} + \Delta E$.  According to Alice, this neutrino can undergo $\nu_\mu \rightarrow \nu_\mu  + e^+ e^-$.

Now consider an observer, Bob,  moving with  respect to Alice in the
same direction as the neutrino, who sees the neutrino to have a
lower energy than Alice.  Such observers exist for which, $E_{Bob} <
E_{thresh} $.  If Bob were to use the same dispersion relation, and
hence the same formula for a threshold, he would conclude that the
neutrino does not emit an electron-positron pair.  There is clearly
a contradiction here.   There are two ways to resolve it.  Either
there is a preferred frame in which case only that preferred
observer, Alice, correctly computes the threshold energy.  Or, if we
insist that the energy-momentum relation is observer independent,
and that both observers agree on the  physics, there cannot be such
an observer-dependent threshold.

As a consequence of this general argument  the framework of deformed
Lorentz symmetry is immune from such anomalous thresholds.


This suffices to conclude in full generality that the concerns for neutrino
superluminality reported in
Refs.~\cite{glashowOPERA,gonzaOPERA,Bi,nussinov}
  do not apply to the case of deformed Lorentz symmetry.
We now illustrate this general result with an explicit calculation.
For this purpose we consider a particularly simple toy model with only neutrinos and electrons/positrons,
such that neutrinos are affected by a deformation of the form
\begin{equation}
 E^2 = p^2 + m_\nu^2 + \frac{2 E^2 p^2}{\mu^2} ~,
 \label{dispGAC}
 \end{equation}
 whereas electrons
 are ordinarily special relativistic.

From Eq.~(\ref{dispGAC})
 it follows that the speed of ultrarelativistic neutrinos (\ref{velmod})
 is given by
\begin{equation}
v  = 1 - \frac{m_\nu^2}{2p^2} + \frac{3 p^2}{\mu^2} ~,
 \label{veloGAC}
 \end{equation}
 and evidently this gives $v > 1$ whenever $p > \sqrt{m_\nu \mu}/6^{1/4}$.

We stress that we consider this scenario
only to illustrate the argument that the
concerns reported in
Refs.~\cite{glashowOPERA,gonzaOPERA,Bi,nussinov} are automatically evaded by any
theory with deformed Lorentz invariance.
It is however intriguing that for $\mu^2=mM_{Planck}$, with $m^2= \Delta m^2 $,
one of the neutrino mass differences, this simple scenario
produces estimates compatible with the OPERA data\cite{JoaoOPERA,lsnote}
(however it does not easily fit the SN1987a data).

The dispersion relation (\ref{dispGAC}) is evidently not  invariant under ordinary,
 undeformed Lorentz boosts. But it  is invariant,
 \begin{eqnarray}
\left[{\cal N}_j^{(\mu)}, E^2 - p^2 - \frac{2 E^2 p^2}{\mu^2}
\right] = 0 ~,
 \nonumber
 \end{eqnarray}
 under the action of deformed boosts with
 the following generators ${\cal N}_j^{(\mu)} $
\begin{eqnarray}
& \delta_j E \equiv [ {\cal N}_j^{(\mu)}, E ] = p_j +  \frac{p^2
p_j}{\mu^2} +  \frac{2 E^2 p_j}{\mu^2} ~,
 \label{boostGACa} \\
& \delta_j p_k \equiv [{\cal N}_j^{(\mu)}, p_k ] = \left(E  -
\frac{p^2 E}{\mu^2} \right) \delta_{jk} ~.
 \label{boostGACb}
 \end{eqnarray}

The final step is to introduce conservation laws that are covariant
under the action of these boosts on neutrino momenta, and the action of standard undeformed boosts
on electron/positron momenta. Since we choose as illustrative case
the process $\nu_\mu \rightarrow \nu_\mu  + e^+  + e^-$ analyzed by Cohen and Glashow,
we explicitly show  a suitable conservation law for that process:
\begin{eqnarray}
{\vec p}&=&{\vec p}' + {\vec k}_{-}+ {\vec k}_{+}
 \label{consGACa}
 \end{eqnarray}
\begin{eqnarray}
E = E' + \Omega_+ + \Omega_- + \alpha \left( \frac{E p^2}{\mu^2}   -
\frac{E' p'^2}{\mu^2} \right)
 ~.
 \label{consGACb}
 \end{eqnarray}
where  ${\vec p},E$ (resp ${\vec p}',E'$) are for the incoming (resp
outgoing) neutrino whereas ${\vec k}_{\pm},\Omega_{\pm}$ are for the
outgoing electron-positron. The parameter $\alpha$ has been
introduced so that $\alpha =1$ for the deformed symmetry framework
and $\alpha =0$ for the case of
 broken Lorentz symmetry.

 It is easy to verify, using (\ref{boostGACa}), (\ref{boostGACb}), (\ref{consGACa}), (\ref{consGACb})
 for the neutrino momenta (and undeformed boosts for electron/positron momenta),
 that our conservation laws, for $\alpha =1$, are covariant.

In order to see that
the process $\nu_\mu \rightarrow \nu_\mu  + e^+  + e^-$ is allowed by broken Lorentz symmetry
but is forbidden by deformed Lorentz symmetry, it is convenient to consider the total
momentum of the outgoing electron-positron pair:
$\Omega_T \equiv \Omega_+ + \Omega_-$ and ${\vec k}_T = {\vec k}_+ + {\vec k}_-$.
Since in our illustrative toy model electrons/positrons are ordinarily special relativistic
also ${\vec k}_T , \Omega_T$ is on a special-relativistic shell:
$M_T^2= \Omega_T^2 - {\vec k}_T^2$. And while the rest energy of the pair is of course
context dependent, for any given electron-positron pair it will be an invariant
and it will be strictly positive: $M_T^2 \geq (2 m_e)^2 >0$.

In terms of ${\vec k}_T , \Omega_T$ and $M_T$
one can use the conservation laws (\ref{consGACa}),(\ref{consGACb})
to obtain the following
\begin{eqnarray}
&p^2= p'^2 +k_T^2 + 2 p' k_T \cos\theta_*
 \label{bombiA}\\
& E = E' + \Omega_T + \alpha \left( \frac{E p^2}{\mu^2}   -
\frac{E' p'^2}{\mu^2} \right)
 ~,
 \label{bombiB}
 \end{eqnarray}
which also involve the opening
 angle $\theta$ between the outgoing-neutrino spatial momentum ${\vec p}'$
 and the total spatial momentum of the outgoing electron-positron pair.

Next we combine these two equations and use dispersion relations to
arrive (also assuming for simplicity that the particles involved are
ultrarelativistic) at
\begin{equation}\cos\theta
 =  \frac{2 E' \Omega_T + M_T^2 + 2 (\alpha - 1) \frac{(E'+\Omega_T)^4-E'^4}{\mu^2}
  -2 \alpha \Omega_T \frac{E'^3}{\mu^2} }
 {2 E' \Omega_T - M_T^2
  \frac{E'}{\Omega_T}- m_\nu^2 \frac{\Omega_T}{E'}   - 2 \Omega_T \frac{E'^3}{\mu^2}}
 ~.\end{equation}
 As it is inevitable in light of the general argument we discussed earlier in this Letter,
 the process is indeed forbidden in the deformed-Lorentz-symmetry case: for $\alpha = 1$
 one formally finds that the process requires $\cos\theta >1$, which is not possible for physical
 opening angles. Of course, the same holds in undeformed special relativity:
 also for $\mu \rightarrow \infty$ (the limit where also the neutrinos of our toy model
 become ordinarily special relativistic) one would need $\cos\theta >1$.
 But the process is possible, as already observed by Cohen and Glashow,
 when Lorentz symmetry is broken:
 for finite $\mu$ and $\alpha =0$, there
are combinations of $E'$ and $\Omega_T$ in the physical phase space compatible
with physical opening angles, such that $\cos\theta \leq 1$.

The derivations proceed in analogous way for verifying that
 also the issues for
pion decay raised by by Gonzalez-Mestres \cite{gonzaOPERA}, Bi et al
\cite{Bi}, and Cowsik et al \cite{nussinov} do not apply
to the deformed-Lorentz-symmetry case.

In conclusion,  Refs.~\cite{glashowOPERA,gonzaOPERA,Bi,nussinov}
 claim that the
OPERA result cannot be interpreted in term of superluminal
propagation. In this paper we observed that their conclusions require
a breaking of lorentz invariance by the introduction of a
preferred frame.  In addition we showed that any description of
neutrino superluminality based on a deformation of special relativity that preserves the relativity of inertial frames would automatically evade their  concerns.
The key to this evasion is an interplay between the deformations of the dispersion relations,
lorentz transformations and conservation laws.

In order to illustrate our argument
we made use of a simple, unrealistic
model of particle dependent deformed lorentz symmetry.  We have not addressed the question of whether there is a
physically realistic model with deformed Lorentz transformations
that can explain the data on neutrino speeds\footnote{Neutrinos in
the context of deformed special relativity have been studied in
\cite{joaomp}.}.


To explain the challenges faced by the construction of such a model we
point out that there are again two possible classes of models.
 The first class involves non-linear realizations of the Lorentz transformations
 obtainable by nonlinear
changes of variables in momentum space (see, {\it e.g.}, Refs.~\cite{leedsr}-\cite{sethmajor}).
 In this approach
 the momentum space picture is straightforward,
 but it is not clear how the
 processes are to be interpreted in spacetime.
The toy model we here used to illustrate our observation belongs to
this class.  It exploits, however, the previously unexplored feature
of having deformations of the lorentz transformations that are
particle dependent. In particular only
 the neutrino dispersion relations and transformation laws need be deformed in relation
 to the OPERA anomaly.  The associated spacetime picture should formally
allow for neutrinos to live in a slightly different spacetime than
the other particles.  Interpreting $v$ in eqs.~(\ref{velmod}) and
(\ref{veloGAC}) as a velocity and insisting also on the relativity
of inertial frames one has to face the apparent non-locality of
distant events \cite{Schutzhold:2003yp, Hossenfelder:2010tm,
AmelinoCamelia:2010qv, Smolin:2010mx}. This may be tolerable in the
neutrino sector, since very little has been established
experimentally about the spacetime localization of neutrinos.

 The second approach and class of models goes under the
 name of {\it relative locality framework} \cite{principle,grf2nd}.  This new principle is realized in
 a class of models in which the deformations arise from modifications of
 the geometry of momentum space such as curvature.
Also in these models different particles would appear to live in different spacetimes, because $x$ and $t$
 are conjugate to  coordinates on a curved momentum space. In this case there is also a large class of curved
 momentum space diffeomorphisms transformations, and it turns out that apparent non-localities affecting processes distant
 from an observer (characteristic of scenarios with particles ``living in different spacetimes")
 are gauge artifacts. Gauge invariant quantities corresponding to physical arrival times can be
computed and reproduce the velocities (\ref{velmod}) and
(\ref{veloGAC}) \cite{Freidel:2011mt, AmelinoCamelia:2011nt}. In
such a relative locality model, one could hope to provide a
comprehensive explanation of the OPERA anomaly without
 running into the  problems raised by Refs.~\cite{glashowOPERA,gonzaOPERA,Bi,nussinov}.

 This remains a task for the future.
 What we have shown here is that deformed Lorentz symmetry provides a large class of counter examples to the
 issues raised in ~\cite{glashowOPERA,gonzaOPERA,Bi,nussinov}.
 If the OPERA results are confirmed, theorists will face a
formidable challenge (see
Refs.~\cite{whataboutopera,JoaoOPERA,op1,op2,op3,op4,op5,op6} for
some of the attempts to deal with this challenge.)  These challenges
 include
 developing quantum field theories consistent with relative locality;
 efforts to do so are underway. What we have
observed in this paper is that the challenge of understanding the OPERA results-if they are confirmed-is more likely to succeed wthin the
framework of deformed rather than broken Lorentz invariance.

\section*{ACKNOWLEDGEMENTS}

We are grateful to many members of the Perimeter Institute community
for their insights, comments and encouragement and also thank Andrew
Cohen for correspondence. GAC and JKG thank Perimeter Institute for
hospitality during their visits in September 2011.  The work of JKG was supported in
part by grants 182/N-QGG/2008/0 and 2011/01/B/ST2/03354.
  Research at Perimeter Institute
for Theoretical Physics is supported in part by the Government of
Canada through NSERC and by the Province of Ontario through MRI.



\begin{thebibliography}{99}

\bibitem{opera}
T.~Adam {\it et al},
arXiv:1109.4897

\bibitem{glashowOPERA}
A.G.~Cohen and S.L.~Glashow ,
arXiv:1109.6562, Phys.~Rev. Lett.~107 (2011) 181803

\bibitem{gonzaOPERA} L. Gonzalez-Mestres,
arXiv:1109.6630

\bibitem{Bi}
  X.-J.~Bi, P.-F.~Yin, Z.-H.~Yu and Q.~Yuan,
  arXiv:1109.6667, Phys.~Rev.~Lett.\ {\bf 107}  (2011) 241802

\bibitem{nussinov}
  R.~Cowsik, S.~Nussinov and U.~Sarkar,
arXiv:1110.0241,  Phys.\ Rev.\ Lett.\  {\bf 107} (2011) 251801.

  \bibitem{N} Z.G.T.~Guiragossian, G.B.~Rothbart, M.R.~Yearian, R.~Gearhart and J.J.~Murray,
  Phys.\ Rev.\ Lett.\  {\bf 34 } (1975)  335.

\bibitem{N+1} A.A.~Abdo {\it et al}, Nature {\bf 462} (2009) 331.  

\bibitem{whataboutopera}
  G.~Amelino-Camelia, G.~Gubitosi, N.~Loret, F.~Mercati, G.~Rosati, P.~Lipari,
arXiv:1109.5172, Int.~J.~Mod.~Phys.~{\bf D20} (2011) 2623.

\bibitem{kosteSMEfirst}
  D.~Colladay, V.~A.~Kostelecky,
hep-ph/9703464,  Phys.\ Rev.\  {\bf D55}  (1997) 6760


\bibitem{colglafirst}
S.R. Coleman and S.L. Glashow,
hep-ph/9703240,
Phys.Lett. {\bf B405} (1997) 249

\bibitem{grbgac} G. Amelino-Camelia, J. Ellis, N.E. Mavromatos, D.V. Nanopoulos and S. Sarkar,
astro-ph/9712103,
Nature {\bf 393} (1998) 763

\bibitem{gacdsr}
G.~Amelino-Camelia,  
hep-th/0012238,  Phys.\ Lett.\ {\bf B510} (2001)  255;
gr-qc/0012051, Int.~J.~Mod.~Phys.~{\bf D11} (2002)  35

\bibitem{jurekdsr}
 J.~Kowalski-Glikman,
hep-th/0102098, Phys.~Lett.~{\bf A286} (2001) 391.

\bibitem{leedsr}
  J.~Magueijo, L.~Smolin,
   hep-th/0112090,
  Phys.\ Rev.\ Lett.\  {\bf 88 } (2002)  190403;
  gr-qc/0207085,  Phys.~Rev.~{\bf D67} (2003) 044017.

\bibitem{judesvisser} S. Judes and M. Visser,
Phys.~Rev.~{\bf D68} (2003) 045001

\bibitem{gacnewjourn} G.~Amelino-Camelia,
 gr-qc/0212002, New J.Phys. {\bf 6} (2004) 188

\bibitem{sethmajor}
D. Heyman, F. Hinteleitner, and S. Major,
gr-qc/0312089,
Phys.~Rev.~{\bf D69} (2004) 105016

\bibitem{JoaoOPERA} J.~Magueijo,
arXiv:1109.6055.

\bibitem{lsnote} L. Smolin,
{\it Note on the OPERA neutrino speed data and relative locality}, unpublished, Sept 25, 2011.

\bibitem{joaomp} M.~Blasone, J.~Magueijo and P.~Pires-Pacheco, Europhys.~Lett.~{\bf 70} (2005) 600;
Braz.~J.~Phys.~{\bf 35} (2005) 447-454.

\bibitem{principle}
  G.~Amelino-Camelia, L.~Freidel, J.~Kowalski-Glikman
and L.~Smolin,
 arXiv:1101.0931
 Phys.~Rev.~{\bf D84} (2011) 084010.

\bibitem{grf2nd}   G.~Amelino-Camelia, L.~Freidel, J.~Kowalski-Glikman
and L.~Smolin,
 arXiv:1106.0313, Gen.~Relativ.~Gravit.~{\bf 43} (2011) 2547.

\bibitem{Schutzhold:2003yp}
  R.~Schutzhold and W.~G.~Unruh,
 gr-qc/0308049, JETP Lett.~{\bf 78} (2003) 431.

\bibitem{Hossenfelder:2010tm}
  S.~Hossenfelder,
  arXiv:1004.0418,
  Phys.\ Rev.\ Lett.\  {\bf 104} (2010) 140402

\bibitem{AmelinoCamelia:2010qv}
  G.~Amelino-Camelia, M.~Matassa, F.~Mercati and G.~Rosati,
 arXiv:1006.2126,
  Phys.\ Rev.\ Lett.\  {\bf 106} (2011) 071301

\bibitem{Smolin:2010mx}
  L.~Smolin,
  arXiv:1007.0718.

\bibitem{Freidel:2011mt}
  L.~Freidel and L.~Smolin,
  arXiv:1103.5626.

\bibitem{AmelinoCamelia:2011nt}
  G.~Amelino-Camelia, M.~Arzano, J.~Kowalski-Glikman, G.~Rosati and G.~Trevisan,
  arXiv:1107.1724, Class.~Quant.~Grav.~(2012, in press)

\bibitem{op1} G.~Cacciapaglia, A.~Deandrea and L.~Panizzi,
arXiv:1109.4980, JHEP {\bf 11} (2011) 137

\bibitem{op2}
G.~F.~Giudice, S.~Sibiryakov and A.~Strumia,
 arXiv:1109.5682

\bibitem{op3} J.~Alexandre, J.~Ellis and N.E.~Mavromatos,
 arXiv:1109.6296

\bibitem{op4} C. Pfeifer and M.N.R. Wohlfarth,
arXiv:1109.6005

\bibitem{op5}
A.~Drago, I.~Masina, G.~Pagliara and R.~Tripiccione,
 arXiv:1109.5917

\bibitem{op6} G.~Dvali and A.~Vikman
arXiv:1109.5685


\end{thebibliography}
\end{document}